# Temperature Dependence of Linked Gap and Surface State Evolution in the Mixed Valent Topological Insulator $SmB_6$

J. D. Denlinger[1], J. W. Allen[2], J.-S. Kang[3], K. Sun[2],
J.-W. Kim[4], J. H. Shim[4,5], B. I. Min[4], Dae-Jeong Kim[6], Z. Fisk[6]

[1]*Advanced Light Source, Lawrence Berkeley Laboratory, Berkeley, CA 94720, USA*
[2] *Dept. of Physics, Randall Laboratory, University of Michigan, Ann Arbor, MI 48109, USA*
[3]*Department of Physics, The Catholic University of Korea, Bucheon 420-743, Korea*
[4]*Department of Physics, POSTECH, Pohang 790-784, Korea*
[5]*Department of Chemistry, POSTECH, Pohang 790-784, Korea*
[6]*Dept. of Physics and Astronomy, University of California at Irvine, Irvine, CA 92697, USA*

**Taken together and viewed holistically, recent theory[1-3], low temperature (T) transport[4,5], photoelectron spectroscopy[6-8] and quantum oscillation[9] experiments have built a very strong case that the paradigmatic mixed valence insulator $SmB_6$ is currently unique as a three-dimensional strongly correlated topological insulator (TI). As such, its many-body T-dependent bulk gap brings an extra richness to the physics beyond that of the weakly correlated TI materials. How will the robust, symmetry-protected TI surface states evolve as the gap closes with increasing T? For $SmB_6$ exploiting this opportunity first requires resolution of other important gap-related issues, its origin, its magnitude, its T-dependence and its role in bulk transport. In this paper we report detailed T-dependent angle resolved photoemission spectroscopy (ARPES) measurements that answer all these questions in a unified way**.

A TI material is a bulk insulator with metallic surface states that are both required and protected by the topology/symmetry of the quantum states that form the bulk insulating gap[10-12]. Understanding the bulk gap is then an essential part of understanding the TI state itself and for $SmB_6$ there are several outstanding issues. First, the TI theory rests[13] on the validity of a Fermi liquid picture[14,15] in which hybridization between Sm 4$f$ and 5$d$ states leads to a gap with an $f$-based X-point conduction band (CB) minimum[16] that is thus far unverified. Second the "large" bulk gap of >14 meV observed in low T ARPES[6-8,17] is much larger than the "small" gap implied by a 3.5 meV activation energy observed in the T-dependent conductivity[18-20], leading to uncertainty as to the actual size of the gap, whether the gap has even been observed in ARPES[21], and the mechanisms of the bulk transport. By performing T-dependent ARPES we directly observe the X-point CB state as it becomes thermally occupied at essentially the "small gap" transport activation energy above $E_F$. With increasing T it shifts downward across $E_F$ and in concert with an extreme broadening of the Sm 4$f$ valence band edge, the "large" hybridization gap is closed. This behavior can then be correlated with the T-dependence of the bulk resistivity, a sign change of the Hall coefficient, and the yet different T-dependence of the bulk valence. Calculated ARPES spectra combining density functional theory with dynamic mean field theory (DFT + DMFT) give important insights into the observed T-dependence of the bulk electronic structure. Of great importance for future TI theory, the in-gap 2D surface states do not simply "disappear"[6-8] upon closure of the hybridization gap, but rather are observed to display a gradual transition to a bulk 3$d$ character.

The temperature-dependent resistivity and Hall transport profiles of $SmB_6$ exhibit three basic regimes of (i) a poor metal at high temperature, (ii) a gradual metal-to-insulator transition (MIT) below 50K with a more than a 5 order of magnitude change in resistivity and (iii) a low temperature residual conductivity regime below 4K[18,19]. The onset of the MIT regime at ≈50K correlates to a sign change of the Hall coefficient[18], a peak in the magnetic susceptibility[22] and a rapid change in the bulk Sm valence[23]. A summary plot of various transport measurements, provided in the Supplemental Online Material (SOM), shows evidence for additional sub-regimes that are demarcated by slope changes in the conductivity, Hall coefficient and bulk valence. For example the thermally activated "small gap" behaviour in the low T part of the MIT regime is more complex than simple Arrhenius in that the activation energy changes from ≈5.6 meV to ≈3.5 meV[20] as T decreases below 25K (Fig. 2a below). The ARPES results presented here over the range of 6K to 190K provide a detailed picture of the T-dependent evolution of the bulk electronic structure in the high T and MIT regimes, as well as that of the 2D in-gap TI surface states that are the origin of the low T residual conductivity.

Figure 1a,b shows for 6.2K an overview of the high symmetry Γ-X-M bulk plane spectra excited at a photon energy of 70 eV. As reported previously, the momentum ($k$) map at $E_F$ in Fig. 1a exhibits four ellipsoidal contours centered on the X-points of the Brillouin zone (BZ)[6-8,21], corresponding to in-gap electron-like surface states[6-8] that propagate up to $E_F$ from Sm 4$f$ states that we observe to disperse between ≈ 14-21 meV below $E_F$. Newly reported here, the four high intensity points ($H$) in the -5 meV map in Fig. 1a along the Γ-M direction correspond to the spectral intensity tails of the Sm 4$f$ states, which make their closest approach to $E_F$ at $H$. Fig. 1b shows various energy dispersions from very near the X-point (X') through the $H$-point, corresponding to the dashed line in Fig. 1a, highlighting the in-gap states, the high-intensity $H$-point dispersion, as well as the bulk X-point Sm 5$d$ electron bands that disperse from below up to the $f$-states. The $k$-location of the $H$-point along the Γ-M direction is coincident with a recently claimed metallic surface state band[24]. As discussed further elsewhere[25] Figure 1a,b clearly shows the $H$-point to be gapped at low T. We note that small surface state pockets observed at Γ[6-8] do not appear for the linear horizontal photon polarization used here. While the existence of those states are important for the topological theory of SmB$_6$[2,3], they are not of interest for the X-point CB focus of this paper.

The X'-$H$-X' momentum cut was continuously monitored as the sample was slowly warmed in few degree steps from 6K to 180K resulting in a 3D data set that is provided as a 44 frame movie in the SOM. Fig. 1b shows ten key temperatures from this data set, with the spectra divided by the T-dependent Fermi-Dirac distribution in order to highlight the spectral weight that is increasingly accessible up to $\approx 4k_BT$ above $E_F$ as T increases. Energies above the 4kT accessibility range have been blanked out in Fig. 1b and 1e. The most striking feature is a distinct movement from 40 K to 78 K of spectral weight within the X-point region, CB states initially above $E_F$ moving to below. A second notable feature with increasing T is a rapid energy-width broadening of the 4$f$ states. An associated dramatic 4$f$ peak amplitude decay, not seen in Fig. 1b due to the use there of individually scaled color tables, is revealed in T-dependent line spectrum stack plots (Fig. 1c) and images (Fig. 1d) of the $H$ and X' points.

The distinct movement of CB states across $E_F$ in the X-point region shows up in the raw data as a spectral hump at $E_F$ (Fig. 1c) and an abrupt appearance of $E_F$ spectral intensity at $\approx$40K (Fig. 1d). In contrast, the $H$-point exhibits only a gradual build up of $E_F$ spectral intensity related to the high temperature broadening of the 4$f$ peak. The 4$f$ energy width broadening and linked amplitude variation reflect changes in the 4$f$ coherence. The amplitude saturation below 15K in Fig. 1c indicates that the 4$f$ states have achieved nearly full coherence. A fuller quantification of the 4$f$ amplitude, energy, width and area is provided in the SOM.

Fig. 1e shows and quantifies all the essential features of the bulk gap. Dividing the Fig. 1d X' image by the Fermi-Dirac distribution reveals quite distinctly the X-point CB state at the point of its first appearance in the $4k_BT$ window $\approx$ 9 meV above $E_F$ at 25K. As T increases and exceeds 100K this band moves to $\approx$ 6 meV below $E_F$ to meet and merge with the rapidly broadening 4$f$ states.

We now relate these findings to the major features of the T-dependent transport. Fig. 2a quantifies the T-dependent downward movement of the CB (labeled "$E_F$ gap") to pass through $E_F$ at $\approx$60K. We see in Fig. 2b that the resulting rapid rise of $E_F$ intensity from 20-50K at the X' point correlates well with the rapid increase of the bulk conductivity, whereas the more gradual increase in $H$-point $E_F$-spectral weight (Fig. 1d) resulting from 4$f$ width broadening does not. The sign change of the Hall coefficient, also plotted in Fig. 2a, correlates well with the X-point CB states passing through $E_F$. Thus the transition to the metallic state is more complex than simple thermal activation to a fixed energy level or the symmetrical[26] closing of a gap around $E_F$. Also it is quite distinct from the closing of the large *f-d* hybridization gap above 100K, estimated and plotted in Fig. 2a (labeled *f*-gap) as the energy difference between the CB energy centroid and the high energy half-width of the 4*f* peak. It is also noted in Fig. 2a that the downwards energy movement of the X-point CB state is greater than the shift to higher binding energy of the occupied 4*f* states, resulting in a net high T narrowing of the overall *f*-bandwidth at the X-point. We interpret this as evidence for a weakening of both *f*-dispersion and *f-d* hybridization as *f*-quasiparticle coherence is lost as T increases.

We also note that the CB states provide an energy scale of the right general magnitude to account for the lower T "small gap" behavior characterized in Fig. 2a by the dual-peaked activation energy derived from the T-dependent conductivity. It is plausible that the range of the activation peaks, from 3.5 meV at 6K to 5.6 meV at 25K is bracketed by the 9 meV X' CB energy level at 25K and the smaller gap to $E_F$, somewhat away from the X-point, implied by the theoretically predicted W-shaped X-point CB dispersion[16,27] (see also Fig. 3a). Due to the narrow energy range imposed by the Fermi-Dirac distribution profile at the lowest T we cannot probe directly for this 3.5 meV state above $E_F$. But good evidence for such a *W*-shaped structure is provided by the $E_F$ intensity profile at 50K (Fig. 2c), which shows two broader peaks (indicated by arrows) between the sharp surface states, also visible directly (dashed lines of Fig. 2c) in the intermediate temperature EDC images in Fig. 1b.

Recently a model has been proposed[21] in which the (001) surface $E_F$ is pinned high in the conduction band above a tiny bulk hybridization gap that has yet to be observed by low T ARPES and that the "in-gap" states are actually of bulk conduction band origin. The observation of X-point conduction band states above $E_F$, their T-dependent behavior and connection to bulk thermal activation and transport clearly show that a large bulk hybridization gap spanning $E_F$ has indeed been observed in ARPES.

The correlated electronic structure of $SmB_6$ is a challenge for theory. T-dependent DFT+DMFT calculations are presented in Fig. 3a and in the SOM, where the mixed valent rather than Kondo origin of the *f*-quasiparticles is also emphasized. Although these calculations succeed for the first time in accurately giving the general 15-20 meV binding energy scale of the bulk *f*-states, they share with all LDA-based results a basic unresolved discrepancy[25,28] with experiment in predicting *f*-sub-band dispersions that extend close to $E_F$ and result in an uncorrelated hybridization gap of ≈15 meV[16,27], that then becomes only meV-scale (or even less) when including correlation-induced energy renormalization needed to make the binding energy scale correct[3,29]. Nonetheless the DMFT calculations describe well the experimental T-dependences of the X-point *f*-conduction band movement downward through $E_F$ and the strong 4*f* state amplitude, width and energy changes, including a coherence saturation below 15K and nearly complete decoherence above 150K (see quantitative analyses in the SOM). Fig. 3a also shows that the associated X-point band narrowing involves mainly the reduced dispersion and separation between the two same symmetry ($X_7$) *f*-quasiparticle states. A small contribution of B 2*p* character to the CB states indicates that this *f*-dispersion and coupling most likely arises from Sm 4*f*-B 2*p* hybridization. In addition DMFT describes (Fig. 3b) the high temperature incoherent electronic structure of bulk *d*-states crossing $E_F$, which is an important aspect of the T-dependence of the in-gap states that is discussed next.

We turn now to the T-dependent evolution of the in-gap surface states. The T-dependent $E_F$ intensity along X'-*H*-X' (Fig. 4a) shows that the in-gap states persist up to temperatures much higher than the >100K closure of the hybridization gap, or the previously reported ARPES "disappearances" at 30K[7], 110K[6] and 150K[8]. This persistence is accomplished by a gradual 2D-to-3D crossover behavior. In great contrast to the 6K data of Fig. 4b, the states crossing $E_F$ at 300K appear to be just extrapolations of the deeper binding energy bulk *d*-state dispersion with similar band velocities. Especially apparent above 60K after closure of the conductivity gap is an evolution of the surface state Fermi momentum ($k_F$) (Fig. 4a) and velocity ($v_F$) (Fig. 4b) towards the high temperature bulk values. Final evidence of the 2D-to-3D crossover of the state crossing $E_F$ comes from its *k*-dependence perpendicular to the sample surface ($k_z$), as shown for (100) normal-emission photon-energy-dependent *k*-maps in Fig. 4c. A $k_F$ value independent of $k_z$ at low T (vertical streaks, also shown for the X'-H-X' cut in Fig. S5a) is replaced at high T by a strong $k_z$ dependence similar to that observed (at all T) for the bulk 5d bands below the f-states (see Fig. S5b).

Thus the high temperature electronic structure that characterizes the poor metallic behavior of $SmB_6$ is that of X-point d-states that no longer coherently hybridize with the f-states and thus are able to penetrate through the now-localized f-states up $E_F$, resulting in bulk X-point ellipsoidal Fermi surfaces with strong incoherent scattering of the electron-like carriers off of the f-states. The surface state $k_F$-shift and $k_F$-width illustrated by $E_F$ intensity line profiles in Fig. 4d are quantified in the SOM and the $k_F$-shift is shown to follow a gradual T-dependent profile similar to the 4f-amplitude coherence changes. Furthermore the $k_F$-width starts to increase upon closure of the conductivity gap above 50K as the in-gap states begin to scatter off f-states, thereby reducing the lifetime of the states from >30 fsec at 6K to 5.5 fsec at 180K. As discussed carefully in the SOM, the $k_F$-shift during this gradual 2D-to-3D crossover seemingly appears to be correlated with an increasing bulk d-state occupation expected from the decreasing f-state occupation as the bulk valence changes from 2.50 to 2.58 ($n_f$ = 5.50 to 5.42) as measured by x-ray absorption spectroscopy[23].

These new T-dependent data verify the X-point conduction band feature essential to the TI theory, show the origin of the small gap transport activation energy, and reveal the relationship between the T-dependences of the dc conductivity and the bulk gap closure. The 2D in-gap states do not simply maintain a fixed $k_F$ value and then vanish with bulk gap closure, but rather exhibit a gradual 2D to 3D evolution as the f-coherence needed to form the bulk gap breaks down, resulting in the high T bad metal state. Describing all this T-dependent behavior is an opportunity to extend TI theory in a new direction that sharply distinguishes weakly from strongly correlated TI materials like $SmB_6$.

**Methods**

**Experimental Methods.** The ARPES measurements were performed at the MERLIN Beamline 4.0.3 at the Advanced Light Source (ALS) synchrotron, using single-crystal samples of $SmB_6$ that were prepared from an aluminum flux and cleaved and measured in ultra high vacuum better than $8\times10^{-11}$ Torr. The ARPES endstation is equipped with a Scienta R8000 electron energy analyzer and a low temperature 6-axis sample manipulator cooled with an open-cycle He flow cryostat. The sample temperature was varied from 6K to 180K in fine steps. The photon energy hv was set to 70 eV, corresponding to a bulk Γ-plane of the simple cubic Brillouin zone, for the T-dependent measurements using 7 meV energy resolution, and was varied from 30 eV to 100 eV for $k_x$-$k_z$ mapping. At 70 eV the photoionization cross-section for Sm *4f* is very high enabling the detailed study of the *f*-state temperature dependence with weak background contribution and yet still clearly observe the in-gap states. Fermi surface maps and $E_F$ data slices employed at most a ±5 meV integration window for improved statistics.

**Theoretical Calculations.** Dynamical mean field theory was combined with density functional theory (DFT+DMFT)[30,31] using the WIEN2K package[32] which is based on the full-potential linearized augmented plane-wave (FLAPW) band method[33]. The DMFT local self energy was implemented only for Sm *4f* electrons and other orbitals were treated in DFT. To obtain the local self energy, a non-crossing approximation (NCA) impurity solver was used with parameters of U = 7.0 eV and J = 0.83 eV. It was also confirmed that the low T limit of the NCA is consistent with that of the one-crossing approximation (OCA)[34]. The Brillouin zone integration was done with a 17×17×17 mesh, and $R_{MT}$'s of Sm and B were chosen to be 2.50 and 1.50 a.u., respectively. The wave function in the interstitial region was expanded with plane waves up to $R_{MT}K_{max}$ = 7.

## Acknowledgements

Supported by U.S. DOE at the Advanced Light Source (DE-AC02-05CH11231), and at U. Michigan (DE-FG02-07ER46379). K.S. was supported by the U.S. NSF (Grant No. ECCS-1307744). B.I.M., J.W.K., and J.H.S. were supported by the NRF (Grant Nos. 2009-0079947 and 2013M2B2A9A03051257). J.S.K. was supported by the NRF (Grant No. 2011-0022444).

**FIGURES and CAPTIONS**

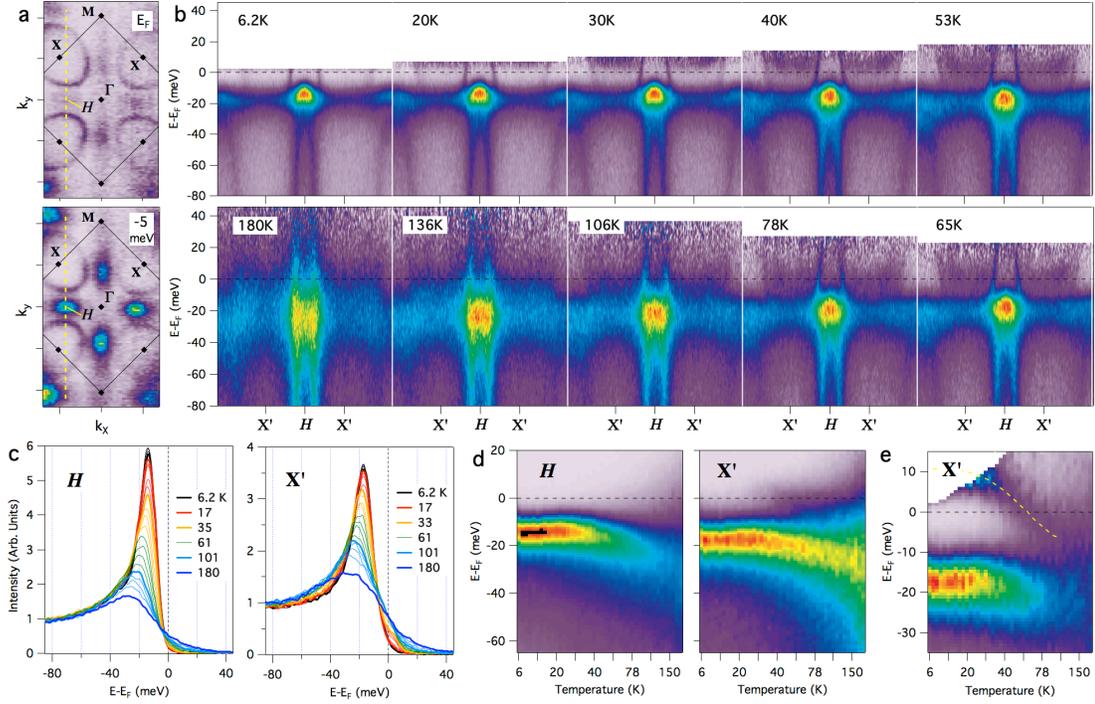

**Figure 1. Temperature-dependent conduction band and 4*f* states of SmB$_6$.** **a**, Constant energy slices at $E_F$ and -5 meV from a k$_x$-k$_y$ angle map at 70 eV and 6K illustrating the vertical <110> sample orientation and the X'-*H*-X' cut used for the temperature-dependent study. **b**, Band dispersion cuts at selected temperatures from 6K to 180K with division by a 4$k_B T$ width Fermi-Dirac distribution to highlight the electronic structure evolution above $E_F$. **c**, Stack plots and **d**, image plots of the energy spectra at the *H* and X' *k*-points highlighting the temperature evolution of the Sm 4*f* amplitude, energy width and energy shifts. **e**, A zoom of the same X'-point data as in **d** with division by the T-dependent Fermi-Dirac distribution to highlight the conduction band movement from above to below $E_F$.

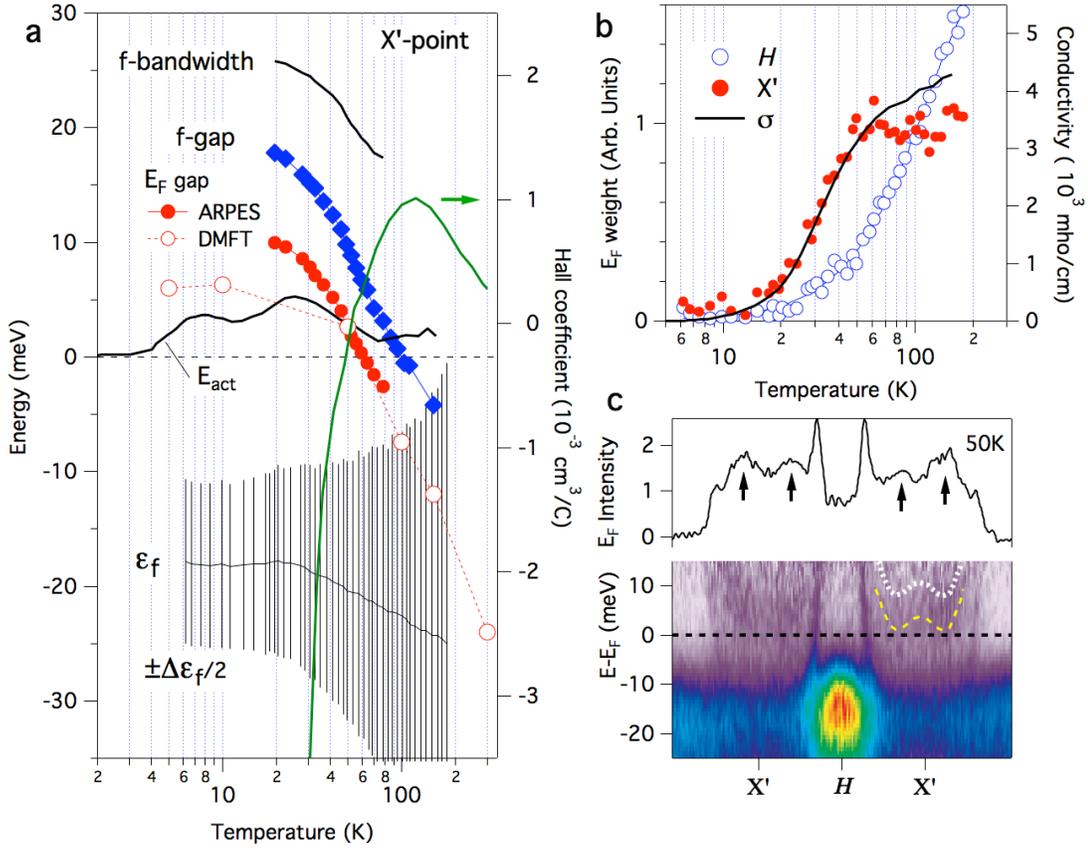

**Figure 2. Correlation of ARPES to bulk transport. a**, Summary of the T-dependent X' conduction band energy movement ($E_F$ gap), the occupied 4f-state energy ($\varepsilon_f$), the f-bandwidth ($E_F$ gap - $\varepsilon_f$) and the difference between the conduction band energy and the 4f-state half-width (f-gap). Comparison is made to an activation energy ($E_{act}$) analysis of the bulk conductivity[20] and to the sign change of the Hall coefficient.[18] **b,** Fermi-edge intensity at the X' and H-points compared to the bulk conductivity[20]. **c**, X'-H-X' EDC at 53K and MDC profile showing sub-structure of states at $E_F$. The dashed line is the theoretically expected shaped band dispersion (see Fig. 3a) which resides above $E_F$ at low T (dotted line).

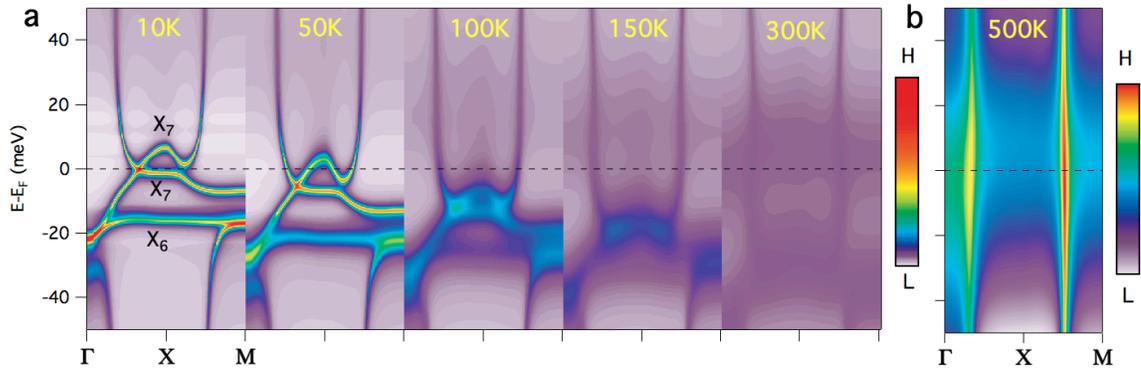

**Figure 3. DMFT calculated temperature-dependent band structure of SmB$_6$.** **a**, Γ-X-M spectral intensity images for T=10K to 300K exhibiting the movement of electronic bands at the X-point from above $E_F$ at low T to below $E_F$ above 50K. A uniform color scale for all temperatures highlights the *4f* amplitude loss and width broadening as *f*-coherence is lost at higher T. **b**, T=500K band image with rescaled color table that highlights the high temperature dispersion of the bulk *d*-bands through the incoherent *f*-states to $E_F$.

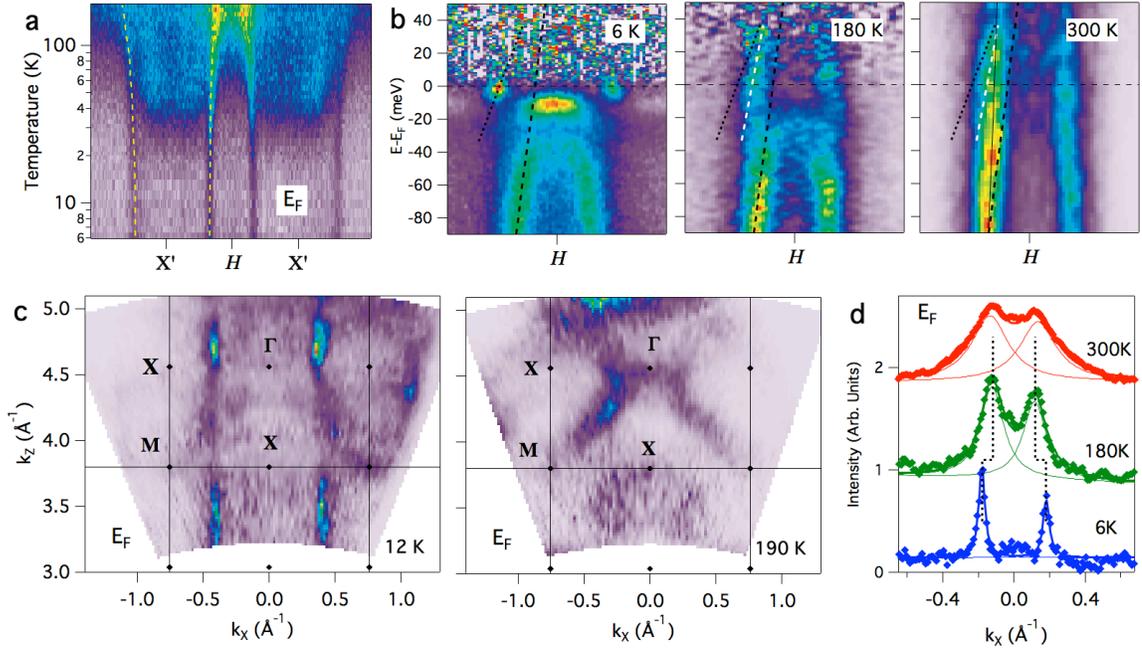

**Figure 4. Temperature dependent evolution of the X-point in-gap states. a**, Fermi-edge intensity slice of the temperature-dependent data set revealing a $k_F$-shift of the in-gap states (increasing X-point electron pocket size) in relation to the rise of $E_F$ spectral weight at the $H$ and X'-point regions. **b**, X'-$H$-X' EDC images at T=6K, 180K and 300K with intensity normalization along the energy axis. Dashed lines assist showing the T evolution of the band velocity and $k_F$ crossing values. **c**, Fermi-edge $k_x$-$k_z$ photon-dependent maps showing 2D surface state behavior at low temperature and 3D $k_z$-dispersive behavior at high temperature. **d**, $E_F$ intensity profiles of the spectra in **b** illustrating the $k_F$ and $k_F$-width changes.

# SUPPLEMENTAL MATERIAL

# Temperature Dependence of Linked Gap and Surface State Evolution in the Mixed Valent Topological Insulator SmB$_6$

J. D. Denlinger[1], J. W. Allen[2], J.-S. Kang[3], K. Sun[2],
J.-W. Kim[4], J.H. Shim[4,5], B. I. Min[4], Dae-Jeong Kim[6], Z. Fisk[6]

[1]Advanced Light Source, Lawrence Berkeley Laboratory, Berkeley, CA 94720, USA
[2] Dept. of Physics, Randall Laboratory, University of Michigan, Ann Arbor, MI 48109, USA
[3]Department of Physics, The Catholic University of Korea, Bucheon 420-743, Korea
[4]Department of Physics, POSTECH, Pohang 790-784, Korea
[5]Department of Chemistry, POSTECH, Pohang 790-784, Korea
[6]Dept. of Physics and Astronomy, University of California at Irvine, Irvine, CA 92697, USA

**Contents**

1. T-dependent transport summary
2. T-dependent X'-H-X' electronic structure movie
3. T-dependent Sm 4*f* quantitative analysis
4. T-dependent in-gap state quantitative analysis
5. Low temperature Two-dimensionality of in-gap states
6. Mixed valency in DMFT

## 1. Temperature dependent Transport Summary

Here we summarize the basic T-dependent transport of SmB$_6$ and discuss the various correlations of different properties and the evidence they give for multiple transport regimes whose boundaries and transitions are relevant to the T-dependent ARPES measurements. Different from the main text presentation, the discussion here is fully integrated with the key T-dependent ARPES results concerning the 4*f* coherence, the inverted *f*-band movement through $E_F$, and the gradual transition of the in-gap states from 2D-to-3D.

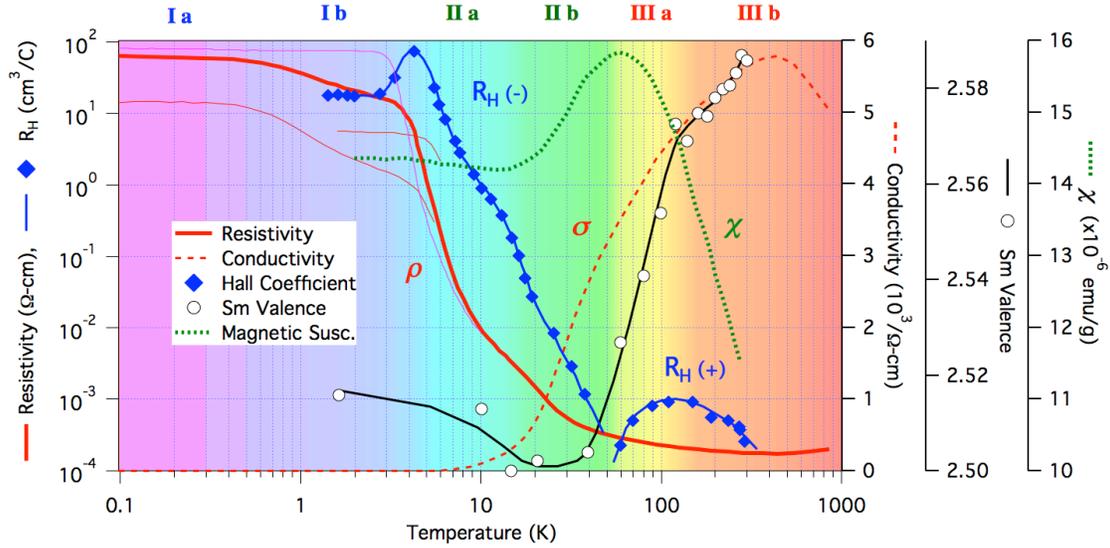

**Figure S1. Temperature regimes of the SmB$_6$ transport and valence**. The three main transport regimes of SmB$_6$ illustrated by comparison of resistivity/conductivity[1], Hall coefficient[2,3] and bulk valence[4] and magnetic susceptibility[5]: (I) low temperature low carrier "in-gap state" regime below 4K, (II) intermediate metallic-to-insulating transition regime with negative Hall coefficient, and (III) high temperature poor-metal regime with positive Hall coefficient above 60K. Multiple low temperature resistivity profiles are plotted to illustrate the variation of the low temperature residual conduction found in the literature. A further subdivision of each regime into two sub-regimes is readily apparent from the various profiles.

Referring to the summary plot in Figure S1, we identify the three basic temperature regimes of the resistivity as (I) a low T residual conductivity regime below 4K, (II) a gradual insulator-to-metal transition from 4K to 50K and (III) a high temperature poor metal regime. Key features that demarcate the transition between regimes I and II are the abrupt rollover of the increasing resistivity and a peak in the negative Hall coefficient. The transition between regimes II and III corresponds to numerous features, (i) a slope change in the conductivity, (ii) a sign change in the Hall coefficient from negative to positive, (iii) a peak in the magnetic susceptibility and (iv) the onset of a rapid change in the bulk valence.

As indicated by the color-coding variations in Fig. S1, various other kinks and slope changes of the plotted properties provide evidence for additional multiple sub-regimes labeled as (a) and (b). We now discuss these sequentially by regime.

**Regime I(a,b)**

The saturated conductivity of regime I is the source of current high scientific interest. Thirty years of difficulties[2,3] with a bulk interpretation have been overcome by the new understanding that it signals transport in protected TI surface states. As referenced already in the main text, strong evidence in support of the TI explanation comes from proofs of the surface origin of the conductivity combined with the literature record of its great robustness, and from spectroscopic (ARPES and dHvA) evidence of 2D metallic states on both the (100) and (110) surfaces. The bold resistivity profile plotted in Fig. S1 represents one of the highest residual resistivities reported in the literature[1] and comes from a float zone grown sample. Other low temperature profiles reported in the literature[6,7] are also plotted in Fig. S1 to illustrate some of the experimental variations in this key property.

In a bulk interpretation the variability of the low temperature residual conductivity has been taken as a clue to the bulk sample quality. Indeed the earliest resistivity measurement[8] reported less than 2 orders of magnitude rise to only 0.02 ohm-cm at low temperature (and also below 4K a Hall coefficient sign change from negative back to positive that was not found in later[2] studies). Knowing now its surface origin means that the strong variations can involve the geometry of the measurement and the ratio of bulk to surface transport, and require a precise knowledge of the sample thickness used to convert the measured resistance to a bulk resistivity.

Low temperature resistivity measurements presented in the literature are predominantly very flat below 4K. However a few of the high resistivity profiles for float-zone grown samples[1,6] exhibit a rising resistivity below 4K (Regime Ib) and true saturation to a flat conduction only below 0.5K (Regime Ia). Thermal activation analysis of the resistivity in Regime Ib gives a very small thermal activation barrier of about 0.1-0.2 meV. The fine details of this regime are below our minimum ARPES temperature of 6K and the residual conductivity is also well below the statistical noise floor for ARPES $E_F$ spectral intensity characterization. Hence we only cite these resistivity variations in Regime I and possibility of two sub-regimes for completeness of discussion and note that fine transport measurements to sub-Kelvin temperatures is a continued and current topic of research[9].

**Regime II(a,b)**

The entire metal-to-insulator transition (MIT) regime from 50K to 4K is commonly described as though it involves a single thermally activated transition with a small gap of about 3.5 meV (or 40K). However it is quite apparent from the resistivity profile and the Hall coefficient that there exists an anomalous bump around 15 K that suggests two different sub-regimes. Thermal activation analysis[6] of the MIT region gives a "small" activation peak of 3.5-5 meV at 7K in Regime IIa before the 15K bump and a slightly larger 6 meV peak at 25K for Regime IIb. The latter value has commonly been multiplied by two assuming a model of states in the middle of a "large" 12 meV semiconductor gap[6]. This dual gap analysis of the Regime II transport is consistent with small and large gaps inferred from early far infrared measurements[10,11].

The fact that activation energy plots derived from the conductivity (see Fig. 2a) exhibit peaks rather than extended regions with a constant value is consistent with the dynamic T-dependent nature of the electronic structure as revealed by the ARPES data in this paper. The ARPES $4f$ T-dependence shows that the $f$-coherence saturates below 15K, coincident with the resistivity and Hall coefficient "bumps" in the middle of Regime II and in between the two thermal activation energy peaks. This suggests the slowing down of dynamic changes in Regime IIb to a more static thermal activation behavior in Regime IIa. As discussed in the main text, the experimentally measured 9 meV X-point conduction band states, combined with a $W$-shaped dispersion can provide a plausible explanation for the low T 3.5 meV to 5.6 meV activation energy range of Regime IIa. This interpretation requires a bulk source of thermally activated electrons at $E_F$. We conjecture this source to be localized bulk defects, consistent with early models of the "small" gap origin[6,10,11] but without the need to invoke a narrow impurity band as done in these early models in order to also explain the residual conductivity below 4K with these same in-gap states. The fact that the $4f$ binding energy of 15-20 meV measured in ARPES is very consistent with zero or only meV level variations, despite different surface preparations of scraping[12], sputter annealed 2×1 reconstruction[13], or (001) cleavage, points to a native defect that is always present to some extent. The origin of low T bulk $E_F$ pinning within the "large" gap is an important topic meriting further study.

Finally, we note that the Hall coefficient exhibits a distinct negative peak at the transition point from rising resistivity in Regime IIa to the flat residual conductivity of Regime I. The origin of this peak has been discussed as being a signature of the crossover between 2D and 3D transport and is reproduced in a simple two-channel bulk and surface conduction model simulation[14].

**Regime II-III transition**

A key temperature dependent ARPES result is that the X-point conduction band begins to move down in energy around 20K when changes in the $f$-coherence also begin and reaches $E_F$ at ≈ 60K, a temperature consistent with the change in the Hall coefficient from negative to positive. The magnetic susceptibility as well as the Knight shift[5] exhibit multiple correlations to the T regimes in Fig. S1. Both measurements exhibit a distinct maximum at 50K, coincident with the regime II-III transition, and both decrease towards lower T before becoming constant below 20K, coincident with the regime IIa-IIb transition. The Knight shift in particular is sensitive to the presence of metallic conduction electrons. Hence the transitional maximum in density of states at $E_F$ likely accounts for the peak in the magnetic susceptibility at 50K that cannot be attributed to that of the localized $Sm^{2+}$ $f^6$ $^7F_J$ multiplets[15] (see Section 6).

**Regime III(a,b)**

A second key temperature dependent ARPES result is that the $f$-coherence is still rapidly changing through the 50K conductivity transition and $4f$ peak amplitude continues to decrease up to and above a coherence temperature of ~150K. Also the hybridization gap, with the stated definition in the main text, "closes" somewhere above 100K and is more closely related to the coherence temperature that is distinctly higher than the conductivity transition at 50K. This coherence temperature scale of 150K divides the high temperature transport into two sub-regimes and is demarcated in Fig. S1 by the two features of (i) a slope change in the bulk Sm valence profile and (ii) a maximum in the positive Hall coefficient. In the scenario that the decoherence of the $4f$ states and weakening of the $f$-$d$ hybridization is mostly complete by 150K, defining the end of Regime IIIa, the slow down in the change of bulk valence would be natural. While a decline in the mobility of holes may contribute to the turnover of the positive Hall coefficient, the subsequent decline towards a compensated metal in Regime IIIb at higher temperature likely reflects the third key ARPES result of an increased contribution of electron-like carriers coming from the heavily scattered X-point electron pockets that emerge from the dimensional crossover of the 2D in-gap states to 3D bulk *d*-states that penetrate the incoherent *f*-states to $E_F$.

## 2. T-dependent X'-*H*-X' Electronic Structure Movie

A supplemental movie of the full temperature dependent data set for the X'-*H*-X' momentum space cut includes 44 temperature steps from 6.2K to 190K, from which only selected E(*k*) images are represented in Fig. 1b. The full data set is represented in the various data slices Fig. 1d, e, and analyses in Fig. 2, 4a, S3 and S4. In addition the movie provides a simultaneous guide to the energy (EDC) and momentum (MDC) slicing of the three-dimensional data set A(ω, k, T) and correlation to the bulk conductivity and *f*-occupation temperature evolution. Figure S2 provides two subsets of the full dataset complementary to Fig. 1 with a uniform color table for all spectra allowing visualization of the decoherence decay of the 4*f* peak amplitude. The larger energy range of Figure S2b also provides a complementary view to Fig. 4b of the T-dependent 5*d* band velocity changes as the *d*-states pass through and hybridize with both $^6H_{7/2}$ and $^6H_{5/2}$ *f*-states and the in-gap states transform from 2D to 3D character.

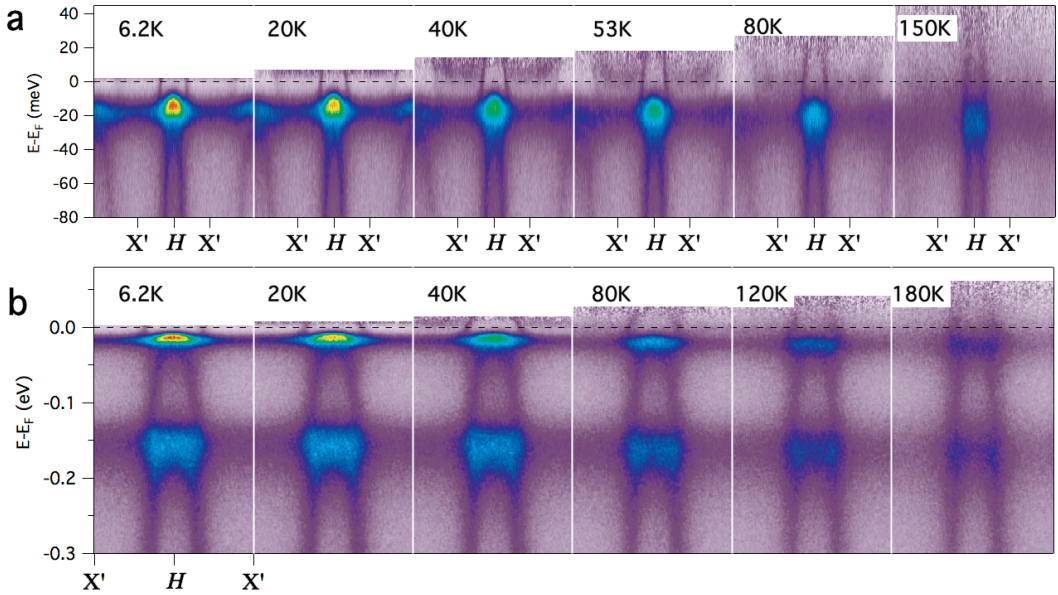

**Figure S2. Temperature-dependent X'-*H*-X' electronic structure of SmB$_6$. a**, Valence band dispersion cuts at select temperatures from 6K to 180K with division by a $4k_BT$ width Fermi-Dirac distribution. The unnormalized raw data illustrates the coherence amplitude decay of the 4*f* states. **b**, 0.3 eV wide spectra illustrating the varying temperature dependence of the *f-d* hybridization effect on the *d*-band dispersion through the two lowest binding energy $^6H_{7/2}$ and $^6H_{5/2}$ final-state multiplets.

## 3. Temperature dependent Sm 4*f* quantitative analysis

Here we compare the quantitative analyses of the temperature dependences of the Sm 4*f* peak at the *H* and X' *k*-points, presented as stack plots in Fig. 1c and images in Fig. 1d. After a Shirley integral background is subtracted, a Gaussian peak fit allows extraction of the peak amplitude, peak area, energy centroid and energy width for each T as shown in Fig. S3.

The **4*f* amplitude variation** is scaled to the height of the previously removed background intensity. First we note that for the linear horizontal (LH) polarization geometry of this measurement, the X'-point has ≈ 4X overall smaller *f*-amplitude and ≈ 2X smaller relative amplitude change than the *H*-point. This difference in 4*f* amplitude between *k*-points likely reflects both the special *f-d* hybridization region of the *H*-point where two bulk *d*-bands from adjacent X-points nearly touch each other as they disperse into the $^6H_{5/2}$ *f*-states, and the reduced number of occupied *f*-states at the X-point, one of which is above $E_F$ at low T.

Nexte we note that previous lower photon energy angle-integrated studies performed at 21.2 eV[12,16] and 8.4 eV[17] photon energies exhibit smaller maximal 4*f*-peak amplitudes of less than 1.7X the high binding energy *d*-state background. The high *f*-state photoionization cross-section at 70 eV in the present measurements allows an unprecedented observation of the full *f*-state coherence changes with an almost negligible background contribution. The dramatic 4*f* amplitude profile decay in Fig. S3 is nevertheless consistent with a previous interpretation of this large 4*f* amplitude variation as a sharpening of the $f$-states below a coherence T of ≈140K[12]. Moreover the fine steps of the amplitude variation show a clear saturation below ≈15K indicating that full 4$f$ coherence is achieved leading into the transport Regime IIa, i.e. that the dynamic coherence-driven electronic structure changes have mostly stopped in this regime characterized by the "small" gap thermal activation of ≈3.5 meV.

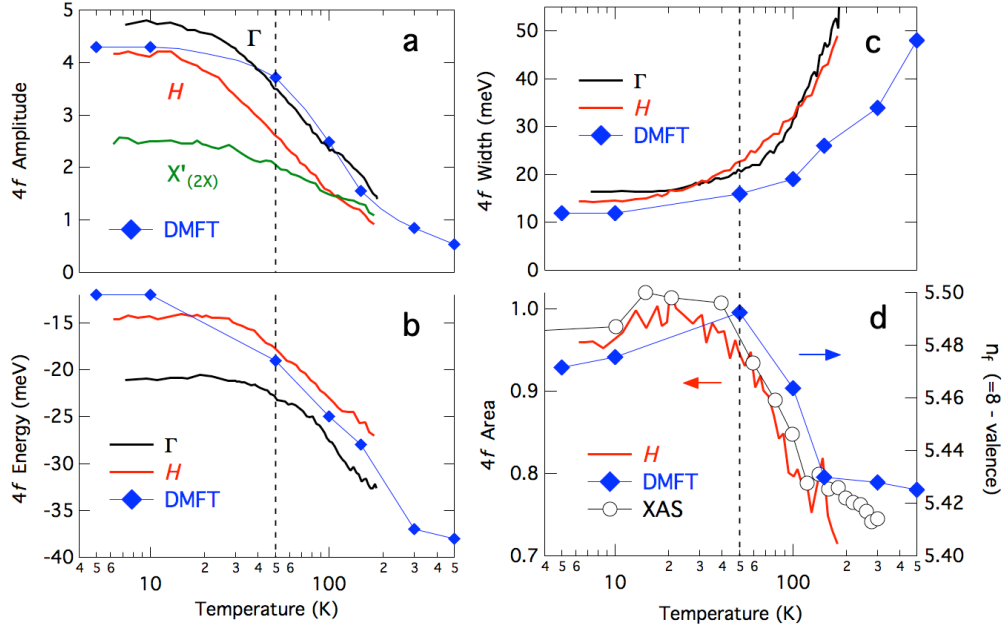

**Figure S3. Quantitative analysis of Sm 4$f$ peak temperature-dependence of SmB$_6$. a**, Sm 4$f$ $^6H_{5/2}$ peak amplitude, **b**, energy centroid, **c**, energy width, and **d**, peak area (=amplitude × width, normalized to 1) at the $H$-point. Comparison is made in each panel to values extracted from $k$-integrated DMFT density of states. Additional comparison is made to the X'-point 4$f$ amplitude, and to the bulk 4$f$-occupation (n$_f$=8-valence) as measured by x-ray absorption.

As the 4$f$ amplitude decreases the **peak width** broadens and the **peak centroid** shifts to higher binding energy. This behavior is interpreted as the system readjusting itself to the changing coherence in order to maintain an approximately constant valence, i.e. the increasing binding energy of the 4$f$ peak compensates for the extreme width broadening that would otherwise push some 4$f$ spectral weight above $E_F$ and result in a decreasing $f$-occupation. The bulk valence as measured by Sm $L_3$-edge x-ray absorption[4] varies from 2.50 to 2.58, i.e. a 1.5% change of $n_f$ (=8 −valence) from 5.50 to 5.42. The $k$-integrated **4$f$ peak area** (≈amplitude × width) for the $H$ and X' points and for a full $k$-integration along

X'-*H*-X' is plotted in Fig. S3 and shows a much less T-dependent decrease as compared to the amplitude. A comparison of these area profiles, scaled to unity at low temperature, to the bulk *f*-occupation is provided in Fig. S3d. In principle with appropriate full *k*-integration over the Brillouin zone as well as integration over all the final state multiplets (not shown in this paper), the experimental 4*f* spectral weight should reflect the bulk valence change as measured by x-ray absorption.

Similar 4*f* peak analyses of DFT+DMFT calculations at 5, 10, 50, 100, 150, 300 and 500K are also plotted in Fig. S3. The 4*f* peak amplitude shows a similar saturation at low temperature and >4X reduction in amplitude up to 150K. The DMFT 4*f* amplitude gives the appearance of a slightly higher coherence temperature scale. The DMFT 4*f* energy shift of ≈15 meV, e.g. from ≈15 meV to ≈30 meV, is a little larger than the ≈10 meV experiment shift of either the *H*- or Γ-points between 6K and 180K. The DMFT 4*f* width also exhibits a slower increase to higher temperatures as compare to ARPES widths that are very similar between *H*- and Γ-points. In contrast to the ARPES experiment, the DMFT calculation enables one to compute the full *k*-integrated density of states over all $Sm^{3+}$ and $Sm^{2+}$ final state multiplets from -25 eV up to $E_F$ and then, by scaling appropriately to the full energy integral (=14 electrons), to extract a theoretical $n_f$ value at each calculated T. The resulting DMFT $n_f$ profile plotted in Fig. S3d shows excellent agreement with the experimental valence variation including a slight decrease from 20K to 4K.

Finally we note that the gradually varying profiles in Fig. S3 do not show any particular correspondence to the 50K rapid conductivity rise or Hall coefficient sign change indicating that the 4*f* state energy-width broadening of spectral weight to $E_F$ resulting from *f*-decoherence is not responsible for the main bulk transport. Rather the main transport changes result from a more complex T-dependent closure to the hybridization gap and associated 4*f* energy shifts, including an overall reduction of the *f*-dispersion band-width at the X-point (discussed in Fig. 3 & S6) that drives the movement of the conduction band states through $E_F$.

## 4. Temperature dependent in-gap state quantitative analysis

Here we discuss the quantitative analysis of the X-point in-gap state temperature dependence from Fig. 4. First we replot in Fig. S4a the 6 -180K $E_F$ intensity image zoomed just around the H-point with comparison to energy slices at -8 meV and -50 meV below $E_F$. The -8 meV energy slice better highlights both the amplitude variation of the spectral tail of the *H*-point *f*-states and provides a clearer view of the increase with increasing T of the X-point in-gap state k-value towards a larger X-point contour size (dotted lines). In contrast the -50 meV energy slice provides a bulk *d*-band reference showing the absence of any significant *k*-shift up to 180K. The in-gap state $k_F$-shift and *k*-width are extracted from Lorentzian peak fits to the spectral intensity profiles as shown in Fig. S4b for both the $E_F$ and -50 meV energy slices for the three selected temperatures of 6K, 180K and 300K. The full T-dependent profiles from the X'-*H*-X' data set plus room temperature (RT) are plotted in Fig. S4c.

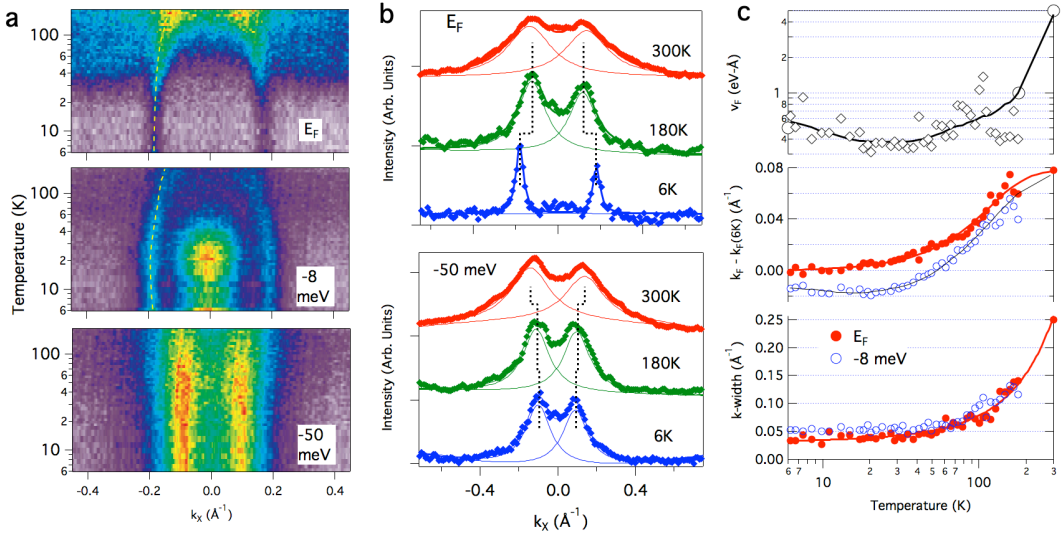

**Figure S4. Quantitative analysis of X-point in-gap state temperature-dependence of SmB$_6$.** **a**, Constant energy slices at $E_F$, -8 meV, and -50 meV of the temperature-dependent data set from 6-180K illustrating spectral weight variation and *k*-shifts. **b**, Example Lorentzian peak fits to the in-gap state spectral intensity profiles at $E_F$ and -50 meV for three select temperatures. **c**, Full T-dependent profiles of the in-gap state *k*-shift, *k*-width, and Fermi velocity $v_F$ (derived from the difference of $E_F$ and -8 meV peak positions).

The $k_F$ **value** of the in-gap states initially begins to change at a low temperature (≈15K) consistent with the 4$f$ coherence profiles in Fig. S3, and increases by up to 0.06 Å$^{-1}$ at 180K before then exhibiting very little additional change up to RT. The increasing size of the X-point 2D contour size of the in-gap state inferred from the $k_F$ shift is seemingly associated with an increasing $d$-state occupation correlated to the decreasing $f$-state occupation as previously discussed with the 4$f$ peak area analysis correlation to the bulk valence changes. However since the 2D in-gap surface states contribute negligibly to the bulk valence properties, details of the 2D-to-3D crossover have to be taken into account for a quantitative comparison of the experimental $k_F$ shifts to the bulk valence. The slow down in $k_F$-shift above 180K can be correlated to the slope change in the bulk valence and maximum in the positive Hall coefficient that divides regimes IIIa and IIIb in Fig. S1. In contrast, the $k$-value of the bulk states at -50 meV changes very little from 6K to 180K, but then exhibits a larger increase up to RT. This is a consequence of increasing band velocity of the $d$-states as the hybridization effect of the $f$-states decreases at higher temperature (e.g. as observed in the RT spectrum in Fig. 4). To quantify the T-dependent band velocity, $v_F$, we also fit the -8 meV energy slice of the data set and use the difference in $k$-values to estimate $v_F$ (=0.008 eV / Δk)  The derived band velocity, plotted in Fig. S4c, exhibits from 6K to 20K an interesting low T velocity decrease in $v_F$ from 0.5 eV-Å to 0.4 eV-Å, before then gradually increasing above 50K to $v_F$ = 1.0 eV-Å at 180K and then rapidly increasing to 5 eV-Å coincident with the bulk $d$-band dispersion at much deeper binding energies (≈1 eV).

A large increase in the **k-width** at high T is apparent in the $E_F$ momentum profiles and is attributable to scattering off now-localized $f$-states. The low temperature -8 meV $k$-width (0.05 Å$^{-1}$) in Fig. S4c is slightly larger than the $E_F$ $k$-width (0.04 Å$^{-1}$) due to being closer in energy to the bulk $f$-states. The $k_F$-width is mostly constant at low temperature up to 50K where the conduction $f$-states just cross $E_F$ and then gradually increases as the hybridization gap continues to close above 100K and the 2D in-gap states transform into 3D bulk states. The $k$-width reflects the imaginary part of the self-energy and the decreasing lifetime of those states at higher T as they increasingly scatter incoherently off localized $f$-states. We can convert the $k_F$-width to a self-energy knowing the band velocity and using the relation, $\Sigma_i = (\Delta k/2)^* v_F$, and then the self-energy to a lifetime using the relation, $\tau$ [fsec] $= \hbar/2\Sigma_i = 330/\Sigma_i$ [meV]. As summarized in Table S1, the self-energy increases dramatically between 180K and RT from both the doubling of the *k*-width and the 5X increase in the band velocity. The corresponding lifetimes decrease from 33 fsec for the 2D in-gap state at 6K down to 5.5 fsec at 180K where the states have mostly evolved into 3D bulk-like states and then further reduces to ≈0.5 fsec at RT.

| Temperature |            | 6 K  | 180 K | 300 K |         |
|:-----------:|:----------:|:----:|:-----:|:-----:|:-------:|
| MDC width   | $\Delta k$ | 0.04 | 0.12  | 0.25  | Å$^{-1}$ |
| Fermi velocity | $v_F$   | 0.5  | 1.0   | 5.0   | eV-Å    |
| *Self-energy* | $\Sigma_i$ | *10* | *60* | *630* | *meV*   |
| *Lifetime*  | $\tau$     | *33* | *5.5* | *0.53*| *fsec*  |
| $k_F$ from X' | $k_F$    | 0.30 | 0.37  | 0.38  | Å$^{-1}$ |
| *Effective mass* | $m^*$ | 4.6  | 2.8   | 0.6   | $m_e$   |
| *Mobility*  | $\mu$      | 13   | 3.5   | 1.5   | cm$^2$/Vs |

**Table S1: Temperature dependent X-point in-gap state properties along X'-*H*.**

The effective mass of the in-gap states can be estimated from the band velocity and the experimental $k_F$ value (relative to X'-point, ≈0.48 Å$^{-1}$) using the relation, $m^* = 7.62\, k_F$ [Å$^{-1}$] / $v_F$ [eV-Å] and the mobility then estimated from the relation, $\mu$ [cm$^2$/Vsec] $= 1.76\, \tau$ [fsec]/$m^*$. The results, summarized in Table S1, show that the 2D in-gap state effective mass is moderately heavy (m*=4.6) at low T, becomes progressively lighter as a 3D bulk state up to 180K and then transforms to a very light $m^*$=0.6, much closer to the bulk LaB$_6$ effective mass of m*=0.5. The 10X reduction in the mobility and 60X reduction in the lifetime between 6K and 300K reflects the hybridization gap closure, and the 2D-to-3D crossover. The high temperature resistivity calculated from the Drude formula, $\rho=1/ne\mu$, using the small 300K mobility of 1.5 cm$^2$/Vs and a carrier density of ≈1 electron/unit cell, ($n=1.4\times10^{22}$ cm$^{-3}$) is consistent with the experimental value of $\rho \approx 2\text{-}3\times10^{-4}$ ohm-cm[1,2] (see Fig. S1).

## 5. Low Temperature Two-Dimensionality of in-gap states

The key ARPES evidence for the X-point in-gap states being two-dimensional at low temperature, and thereby of surface state origin, comes from the lack of $k_z$-dispersion of the in-gap state parallel $k_F$ vector as probed by variation of the photon energy[18-20]. Since the 2D surface state identity (and hybridization gap size) has recently been called into question[21], we supplement our low temperature normal emission X-Γ-X $k_z$-dependence Fermi surface data set in Fig. 4c with an additional 6.2K off-normal photon-dependent cut in Fig. S5 that passes through X'-$H$-X' at 70 eV. Here again, the Fermi-edge map shows very straight lines along $k_z$ inconsistent with a bulk state and in stark contrast to the strong X-point electron-like $k_z$-dispersion of the bulk $d$-states below the $f$-states (-40 meV) that defines the special $H$-point dispersion of the $f$-states.

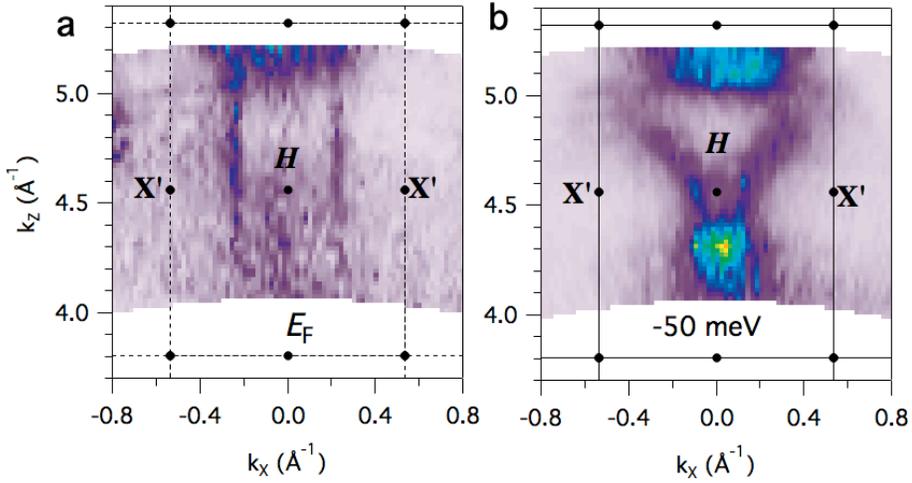

**Figure S5: Low temperature photon-dependent mapping through X'-$H$-X' for SmB$_6$. a**, Fermi-edge intensity map measured at T=7K spanning hv=55-95 eV probing the in-gap states. **b**, energy slice of the same data map at -50 meV probing bulk $d$-states. The in-gap states crossing $E_F$ exhibit straight vertical lines along $k_z$ indicative of 2D surface state origin in contrast to the strong $k_z$-dispersion of the bulk $d$-states.

## 6. Mixed Valency in DMFT

SmB$_6$ is most commonly referred to as a "Kondo" insulator despite the experimental evidence for it being strongly mixed valent, with an average Sm valence of ≈2.5, and thus well away from the Kondo regime near integer valence where virtual charge fluctuations lead to an emergent low energy scale for spin fluctuations. The experimental small 4$f$ binding energy in SmB$_6$ is consistent with strong mixed valence in which real charge (i.e. valence) fluctuations control the physics.

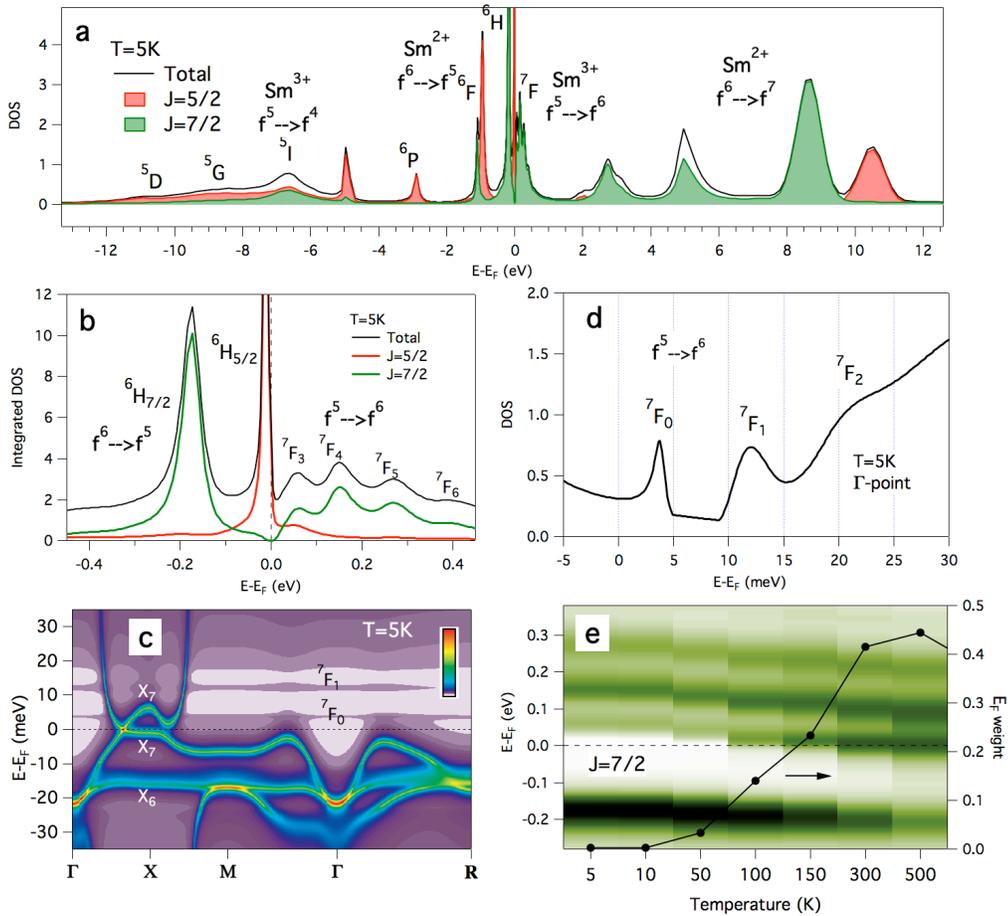

**Figure S6: DMFT calculated mixed valency in SmB$_6$.** **a**, $k$-integrated DMFT density of states (DOS) at 5K showing both Sm$^{3+}$ and Sm$^{2+}$ $f^N \to f^{N-1}$ final state multiplets for $E<E_F$ and $f^N \to f^{N+1}$ final state multiplets for $E>E_F$. **b**, Expanded view of the $k$-integrated multiplets peaks near $E_F$. **c**, $k$-resolved band dispersions at 5K showing constant energy streaks above $E_F$ corresponding to Sm$^{3+}$ $f^5 \to f^6$ $^7F_J$ multiplets. **d**, Expanded view of the lowest energy Γ-point multiplet peaks above $E_F$. **e**, Temperature dependence of the J=7/2 $k$-integrated DOS compared to the X-point $f$-sub-band energy shifts and (right axis) $E_F$ weight profile arising from the $^7F_J$ state energy movement.

Below we explore the mixed valency of $SmB_6$ from the viewpoint of the theoretical DMFT calculation and discuss the effects of valence fluctuations. Fig. 6a shows the wide energy range *k*-integrated DMFT spectral function of $SmB_6$ including not only the final state multiplets of the $Sm^{3+}$ $f^5 \rightarrow f^4$ and $Sm^{2+}$ $f^6 \rightarrow f^5$ one-electron removal processes below $E_F$, but also the one electron addition final state multiplets of the $Sm^{3+}$ $f^5 \rightarrow f^6$ and $Sm^{2+}$ $f^6 \rightarrow f^7$ processes above $E_F$. The expanded energy scale of Fig. S6b shows the excitations near $E_F$, $^6H_{5/2}$ and $^6H_{7/2}$ for $f^6 \rightarrow f^5$ and $^7F_J$ for $f^5 \rightarrow f^6$. The J=0 and J=1 states of the latter can be seen in Fig. S6c and S6d, where they appear at 4 meV and 12 meV, respectively, as low intensity narrow excitations over much of the Brillouin zone. As set forth in early Fermi liquid theory[22,23] the low lying *f*-quasiparticles are formed from the weight of the $^6H_{5/2} \leftrightarrow {}^7F_0$ fluctuations and appear in Fig. S6c below $E_F$ and in the X-point region dispersing above $E_F$ to an energy slightly greater than that of the $^7F_0$ elsewhere in the Brillouin zone. One might expect that similar to the T-dependence of the CB quasiparticle states, the $^7F_J$ weight would also shift down in energy with increasing T and Fig. S6e shows that this is indeed the case. However the buildup of the weight at $E_F$ due to these states is relatively small and with a T dependent profile most similar to the x-ray absorption valence profile and not to the bulk conductivity.